\documentclass[aps,prd,10pt,showpacs]{revtex4-2}
\usepackage{amsfonts}
\usepackage{amsmath}
\usepackage{amssymb}
\usepackage{graphicx}
\usepackage{color}

\usepackage[utf8]{inputenc}
\usepackage{epsfig}
\usepackage{bm}
\usepackage{hyperref}
\usepackage{times}

\def\be{\begin{equation}}
\def\te{\end{equation}}
\def\nn{\nonumber\\}
\def\ba{\end{eqnarray}}
\def\ea{\end{eqnarray}}
\def\bea{\begin{eqnarray}}
\def\tea{\end{eqnarray}}

\begin{document}

\title{On the Energy Spectrum of Non-Newtonian Turbulence}

\author{Esteban Calzetta}
\email{calzetta@df.uba.ar}
\affiliation{Universidad de Buenos Aires, Facultad de Ciencias Exactas y Naturales, Departamento de Física, Buenos Aires, Argentina,\\
and CONICET-Universidad de Buenos Aires, Instituto de Física de Buenos Aires (IFIBA), Buenos Aires, Argentina  }

\begin{abstract}
The goal of this paper is to propose a  theoretical framework to study homogeneous and isotropic turbulence in a viscoelastic fluid, regarded as a perturbation of a Newtonian incompressible fluid, where the fluid relaxation time, or else the Weissenberg number, plays the role of small parameter. We use a Martin-Siggia-Rose framework to obtain a formal expression for the velocity correlation function of the non-Newtonian flow, and we expand this formal expression to linear order in the relaxation time. The coefficients in this expansion are correlation functions of the base Newtonian flow. We do not derive these correlations, instead we replace them by their values according to K41 theory, which could be regarded as an extreme form of renormalization. While substantial work will be necessary to validate the model against numerical and experimental data, preliminary results are encouraging.
\end{abstract}

\maketitle

\section{Introduction}\label{intro}

The goal of this paper is to propose a  theoretical framework for studying turbulent flows in situations beyond K41 theory \cite{Chandra,MonYag71,FRIS95} yet connected to it. The example we discuss is the spectrum of velocity fluctuations in homogeneous, isotropic turbulence of viscoelastic fluids \cite{Lumley64,Seybold19,Zhang21,SZ22,MT21,Ros23,ACR23,Ros24,Ros24b}, which reduce to Newtonian incompressible fluids when the relaxation time is taken to $0$. Another example, closely connected to the first, see Appendix \ref{CM}, is the turbulent flow of a relativistic viscous fluid \cite{ED18,EC1}, which becomes a Non-Newtonian fluid when the speed of light $c\to\infty$.

The basic idea is to formulate the problem within a Martin-Siggia-Rose (MSR) framework \cite{MSR73,Ph77,Eyink,Kamenev,JZEC,DeDGia06}. In this approach, the Schwinger-Dyson equations for the causal response and the velocity correlations are derived from an effective action (EA) \cite{Ram07,CalHu08,Kovtun,KMR,HKR,Haehl18,MGKC}, which has a formal expression as a path integral over velocity and stress tensor fluctuations (details are provided in Section \ref{model}). We then expand this formal expression in powers of the relaxation time (or its dimensionless equivalent, the Weissenberg number). The coefficients in this expansion can be written in terms of correlation functions in the base Newtonian flow. We do not derive these correlations from the MSR EA itself; instead, we substitute them with convenient ansätze suggested by K41 theory. This will allow us to sidestep some known criticisms of the MSR formalism in the literature \cite{K71,GBF81,G98,Tsin09,MC23,MC24}. We  discuss this strategy in the Final Remarks Section \ref{Final}; for the time being, we simply propose it as a matter of expediency.

Concretely, we begin by assuming that the Navier-Stokes equations (NSE) provide the ``bare'' description of homogeneous, isotropic turbulence in a Newtonian incompressible fluid. Then we perturb the NSE by adding non-Newtonian terms in the fluid stress tensor \cite{GS72,DE86,Sar07,MS15}. The perturbation we consider arises in models of polymeric solutions \cite{Lum69,Virk75,Gen90,Bir87,Cal10,Muller22}. In Appendix \ref{CM} we show that this model also describes the non-relativistic limit of a conformal fluid \cite{RZ13,RR19,LCEC}.  We then work out the first order correction to the spectrum from the perturbed EA. We thus find a concrete analytical expression for the velocity correlation function of the viscoelastic flow, see eq. (\ref{spectrum}) below. Validating this expression against numerical and experimental data is a massive undertaking; preliminary comparisons at moderate Reynolds numbers and low Weissenberg numbers, such as the data presented in \cite{ACR23}, are encouraging. 

Before getting  into the details of our proposal let us summarize the main concepts involved.

The most direct way to approach turbulence from the perspective of field theory is to treat turbulence as an ensemble problem. There is an ensemble of many realizations of the flow, and the goal is to extract the statistics of the flow over the ensemble. Correspondingly, in the following all expectation values will mean averages taken over the ensemble. 

In order to achieve a statistically stationary situation in spite of dissipation, each realization of the flow must be forced. We shall assume each realization is forced in a different way, thus there will be a statistical distribution of the external forces over the ensemble. Concretely we shall assume the external forces are Gaussian, white in time and homogeneous and isotropic in space \cite{Novikov1,Novikov}. We shall not adopt a concrete force self-correlation over and above the assumption that forcing is effective mostly in the creation range, so that velocity fluctuations in the inertial and dissipative ranges are due to nonlinear and dissipative processes characteristic of the flow dynamics. The underlying assumption is that turbulence, at least in the inertial and dissipation ranges, is an universal phenomenon, at least in those aspects relevant to the discussion below, so the correlations derived from random forcing are in fact equivalent to those found from other ways to induce turbulence.  This assumption is supported by numerical realizations of the random forcing scenario \cite{ESP88,MD98,Alv99,BCPW19,IDES25,MTBBBM25}.

We shall also assume the net force acting on the whole flow vanishes. This allows us to choose a frame where the total momentum of the flow is zero. Therefore random Galilean symmetry is broken, and there is a one-to-one correspondence between flow and forcing in each realization. We discuss further this issue in Appendix \ref{RGI}.

Our goal is to find the energy spectrum of a non-Newtonian fluid described in eqs. (\ref{eoms}) below, in the limit where the parameter $\tau_1=\tau_2=\tau$ is small. The energy spectrum is related, though the Wiener-Khinchin theorem, to the Fourier transform of the equal time velocity-velocity correlation. 

One may derive this correlation from a generating functional, first introduced in this context by Hopf \cite{Hopf}. To this end one introduces a fictitious external force, to be taken to zero at the end of the calculation. Then the velocity-velocity correlation is a second derivative of the generating functional with respect to this external force. This is similar to finding the magnetization density correlation function of an Ising model from the second derivatives of the free energy with respect to an external magnetic field. 
 
However, instead of finding the free energy as a function of the magnetic field, we may go through a Legendre transform and obtain an equivalent free energy as a function of the  magnetization density itself. In the same way the Legendre transform of the generating functional yields a new functional of the mean velocity field, the so-called Martin - Siggia - Rose effective action (MSR EA), already introduced above. The MSR EA contains all the information about the statistics of the flow, and in this respect it is fully equivalent to the generating functional or to the probability density function of velocity fluctuations.

The important point is that within the random forcing scenario, both the construction of the generating functional and the effective action are some very basic and well understood formal procedures, which do not add nor detract from the physical content of the model, which is defined by the equations of motion (\ref{eoms}). These approaches are also equivalent to Hopf's Fokker-Planck equation for the evolution of the probability density function, which we discuss in Appendix \ref{MSR4LET}. In this sense there may be no reasonable doubt about the validity of the formal expressions obtained from field theory techniques. The problem is how to actually compute them.

In the effective action approach the correlation functions are obtained as the  solution of a differential equation, the so-called Schwinger-Dyson equation. The coefficients in the Schwinger-Dyson equation are in turn derivatives of the effective action with respect to the mean fields, as will be detailed below, see Section \ref{model}. Field theory yields a formal expression for the effective action as a Feynman path integral over all flow configurations in space time \cite{Rivers}. 

We wish to deploy these methods to study non-Newtonian flows which remain close to high Reynolds number Newtonian turbulence, as described by eqs. (\ref{eoms}) below. We may use the path integral representation to obtain an expansion of the derivatives of the effective action in powers of the Weissenberg number, defined below, see eq. (\ref{Wi}), which we adopt as small parameter.  The coefficients in this power expansion may be obtained from correlation functions computed in the base Newtonian flow.

At this point we make a fundamental assumption, namely that Newtonian turbulence, regardless of the forcing mechanism, is well described by K41 theory, at least in those aspects which are relevant to the present computation. This assumption may be disputed, but we believe the supporting evidence, both experimental  and theoretical, is strong enough to take it, at least, as working hypothesis, to be improved upon in the future. 

This assumption allows us to use K41 theory to compute the correlations which make up the Weissenberg number expansion of the effective action, bypassing the problem of actually computing the same directly from the path integral.

In summary, we use the MSR path integral expression for the effective action to find a formal expansion, in powers of the Weissenberg number, of the Schwinger-Dyson equations which yield the velocity - velocity correlation of a non-Newtonian turbulent flow. The coefficients in this expansion are shown to be given by further averages computed in a fiducial Newtonian flow. We then use K41 theory to compute these averages, instead of attempting to get them from the path integral itself.

This strategy resembles the renormalization method in quantum field theory, where sums of Feynman diagrams representing measurable quantities (such as a particle's charge) are replaced by their experimentally measured values in further computations, rather than attempting an actual calculation - which in quantum field theory yields meaningless results \cite{Collins}.

Observe that we use perturbation theory to find the Schwinger-Dyson equations, but then solve them as they are, so the final spectrum eq. (\ref{spectrum}) below displays a complex $\tau$ dependence.

A similar strategy is used in the so-called leading log resummation of a perturbative series, where one seeks exact solutions of renormalization group equations which, themselves, were derived to a finite order in perturbation theory \cite{Collins}.

Let us add a brief comment on our choice of small parameter. The non-Newtonian model we are studying reverts to Newtonian on time scales longer than $\tau$, equal to $\tau_1$ and $\tau_2$ in eqs. (\ref{eoms}). On the other hand, turbulent fluctuations only remain coherent on a time scale of the order of $\lambda^2/\nu$ , where $\lambda$ is a characteristic lenght and $\nu$ is the viscosity. It is natural to choose $\lambda=\lambda_0$ the Taylor microscale. The Taylor microscale is an average scale, where the average is weighted by the energy spectrum; see eq. (\ref{lambda0}) below for a precise definition. The ratio of these two time scales yields a figure of merit, the Weissenberg number Wi in eq. (\ref{Wi}). Wi measures how strongly the non-Newtonian flow departs from Newtonian phenomenology, which is obtained in the Wi$\to 0$ limit. A small Wi means that non-Newtonian effects are transient on the time scales relevant to the base flow. For further discussion see \cite{Poole}.

This paper is organized as follows. Next section \ref{model} presents the basic notations and the equations of the fluid, both the NSE and the perturbed one. To make the work self-contained, we include an introduction to the MSR approach and the EA therefrom. We show how to derive the correlation functions from the EA and develop their perturbative expansion. In this Section we have adopted a highly compressed notation that best displays the structure of the problem, but we return to a more natural notation in the remaining Sections.

In section \ref{PES} we proceed to compute the kernels necessary to write down the perturbed Schwinger-Dyson equations, solve for the causal response function and finally derive the analytical expression for the spectrum of velocity correlations eq. (\ref{spectrum}), which is the main result of this paper. We begin with a brief review of the correlations in K41 theory. While this contains no new material, it is convenient to have all the basic formulae in one place.  

In section \ref{results} we introduce the overall velocity scale $u_1$ (see eq. (\ref{u1})) and the Taylor microscale $\lambda_0$ (see eq. (\ref{lambda0})) \cite{TL}. This allows us to define two dimensionless numbers, Reynolds number and Weissenberg number (see eqs. (\ref{Re}) and (\ref{Wi})) \cite{MS15}, which make it much easier to compare our results with the literature. We conclude with some brief final remarks in section \ref{Final}. 

The paper includes five appendices. In Appendix \ref{CM} we show how the model in Section \ref{model} describes the nonrelativistic limit of a conformal fluid. In Appendix \ref{RGI} we discuss how to account for the random Galilean invariance \cite{Krai64,HorLip79,BH05,BH07,BH09} of the NSE in the EA formalism.  This subject, which we left out of the main text for simplicity, has deep implications for the development of the theory. Appendixes \ref{Lambdacomp}and \ref{spectcomp}  contain some technical details. Appendix \ref{MSR4LET} shows how one may use the MSR technique to derive the evolution equation for the probability density function of a randomly stirred fluid.

\section{The model}\label{model}
The model is represented by the equations

\bea
&&Q^j=V^j_{,t}+V^kV^j_{,k}+P^{jk}_{,k}+\frac1{\mu}P^{,j}=0\nn
&&Q^{jk}=\tau_1\left[P^{jk}_{,t}+V^lP^{jk}_{,l}\right]-\tau_2\left[P^{jl}V^k_{,l}+V^j_{,l}P^{lk}\right]+P^{jk}+\nu\Sigma^{jk}=0
\label{eoms}
\tea

\be 
V^j_{,j}=0
\te
where $V^j$ is the incompresible fluid velocity, $P^{ij}$ is the stress tensor,  $P$ is the pressure, $\mu$ is the constant fluid mass density, $\Sigma^{ij}$ is the shear tensor

\be
\Sigma^{jk}=V^{j,k}+V^{k,j}
\label{shear}
\te
and $\nu$ is the kinematic viscosity. When $\tau_{1,2}\to 0$ at fixed $\nu$ we get an ordinary Newtonian fluid. When $\tau_1=\tau_2=\tau$, the derivative terms in the second of eqs. (\ref{eoms}) add up to the upper convected derivative of $P^{ij}$ \cite{MS15}. When $\tau_2=0$ it reduces to a material derivative, which is the case that describes the nonrelativistic limit of a conformal fluid, see Appendix \ref{CM}. In this note we shall assume $\tau_1=\tau_2=\tau$. 

We note that the right hand side of equations (\ref{eoms}) ought to display stochastic sources necessary to put the fluid in motion. However, since we wish to work in the regime where fluid fluctuations are self-sustained, we shall not consider these sources explicitly.

To be able to derive equations (\ref{eoms}) from a variational principle we introduce Lagrange multipliers $A_{j}$ and $B_{jk}$ such that $A^j_{,j}=0$, and write

\be
S=\int d^3ydt\;\left\{A_jQ^j+B_{jk}Q^{jk}\right\}
\label{af}
\te
We delete the pressure term from $Q^j$, since it integrates to zero anyway.

\subsection{The MSR EA} \label{MSR}

We see that the action functional eq. (\ref{af}) depends on four different fields, the physical fields $V^j$ and $P^{ij}$ and the auxiliary fields $A_j$ and $B_{ij}$. This diversity makes for a rather complex field theory. 

To avoid unnecessary complications, we shall adopt an scheme based on three levels of description. Eqs. (\ref{eoms}) and (\ref{af}) belong to the first level, where we treat both physical and auxiliary fields as distinct. In the second level, however, we drop this distinction and gather together the physical fields into a single string $V^a=\left( V^j,P^{jk}\right) $, and similarly the auxiliary fields into a string $A_a=\left(A_j,B_{jk} \right) $. For higher compression, in the third level of description we regard all variables as components of a single object $X^J=\left( V^a,A_a\right) $. In the second and third levels space-time indexes are included into the indexes $a,J$ and we apply Einstein's convention to sums over indexes, both discrete and continuous.

Given an action $S\left[ X\right] $ we define a generating functional 

\be 
e^{iW\left[ J\right] }=\int DX\;e^{i\left( S\left[ X\right] +J_KX^K\right) }
\label{gf}
\te 
were the $J_K$ are a string of external sources. Differentiation yields the mean fields 

\be 
\bar X^J=\frac{\delta W}{\delta J_J}
\label{s2f}
\te 
We shall work under conditions where symmetry forces all background fields to zero, namely homogeneous, isotropic turbulence. Further differentiation produces the higher cumulants, in particular the two-point correlations

\be 
\frac{\delta^2 W}{\delta J_J\delta J_K}=i\left\langle X^JX^K\right\rangle 
\label{2pc}
\te 
where we are already using the fact that the mean fields vanish. It is convenient to choose the mean fields, rather than the sources, as independent variables. To achieve this, we introduce the effective action $\Gamma$ as the Legendre transform of the generating functional 

\be 
\Gamma\left[ \bar X\right] =W\left[ J\right] -J_K\bar X^K
\label{1PIEA}
\te 
whereby we get the equations of motion for the mean fields

\be 
\frac{\delta \Gamma}{\delta \bar X^J}=-J_J
\label{f2s}
\te  
Differentiating eq. (\ref{s2f}) with respect to the mean fields and using eq. (\ref{2pc}) we get 

\be 
\frac{\delta^2 \Gamma}{\delta \bar X^J\delta \bar X^K}\left\langle X^KX^L\right\rangle =i\delta^L_J
\label{SchDy}
\te
$\delta^L_J$ denotes the identity operator in the corresponding functional space. Similarly, from eq. (\ref{f2s}) we get 

\be 
\left\langle X^JX^K\right\rangle\frac{\delta^2 \Gamma}{\delta \bar X^K\delta \bar X^L} =i\delta^J_L 
\label{SchDy2}
\te 
These are the Schwinger-Dyson equations of the theory. From either of these equations we can derive the two-point correlations from the effective action.

\subsection{Auxiliary and physical fields}

We will now elaborate on the analysis above by distinguishing physical fields $V^a$ from auxiliary fields $A_a$. We also distinguish the external sources $J_a$ coupled to physical fields from the sources $K^a$ coupled to auxiliary fields. The action eq. (\ref{af}) is written as

\be
S=A_aQ^a\left[V\right]
\te 
The equations of motion $Q^a$ are causal and we assume (\cite{Zin93})

\be
{\rm{Det}}\frac{\delta Q^a}{\delta V^b}=\;{\rm{constant}}
\te 
We may choose the constant to be $1$. The generating functional eq. (\ref{gf}) is expanded into

\be
e^{iW\left[J,K\right]}=\int DADV\; e^{i\left(A_aQ^a\left[V\right]+A_aK^a+J_aV^a\right)}
\te 
Observe that

\be
W\left[0,K\right]=0
\te 
identically, so all the expectation values of products of auxiliary fields vanish. Eq. (\ref{SchDy}) becomes

\be
\left(\begin{array}{cc}\Gamma_{,\bar{A}_a\bar{A}_b}&\Gamma_{,\bar{A}_a\bar{V}^b}\\\Gamma_{,\bar{V}^a\bar{A}_b}&\Gamma_{,\bar{V}^a\bar{V}^b}\end{array}\right)\left(\begin{array}{cc}0&\left\langle A_bV^c\right\rangle\\\left\langle V^bA_c\right\rangle&\left\langle V^bV^c\right\rangle\end{array}\right)=i\left(\begin{array}{cc}\delta^a_c&0\\0&\delta^c_a\end{array}\right)
\te
This implies that $\Gamma_{,\bar{A}_a\bar{V}^b}$ and $\left\langle \bar{V}^b\bar{A}_c\right\rangle$ are non singular, since

\be 
\Gamma_{,\bar{A}_a\bar{V}^b}\left\langle V^bA_c\right\rangle=i\delta^a_c
\label{Jeq}
\te
and then it must be 

\be
\Gamma_{,\bar{V}^a\bar{V}^b}=0
\te 
when the mean auxiliary fields vanish.

The correlations of type $\left\langle V^bA_c\right\rangle$ are the response functions of the theory. Once they are found, the physical correlations $\left\langle V^aV^b\right\rangle$ follow from

\be 
\left\langle V^aV^b\right\rangle=i\left\langle V^aA_c\right\rangle\left\langle V^bA_d\right\rangle\Gamma_{,\bar{A}_c\bar{A}_d}
\label{physcors}
\te 
The second derivatives $\Gamma_{,\bar{A}_c\bar{A}_d}$ are the so-called ``noise kernels'' \cite{CalHu08}

\subsection{Computing the EA with the background field method}\label{BFM}

According to the usual rule \cite{CalHu08}, the EA is the classical action plus a ``quantum correction''

\be
\Gamma=S+\Gamma_Q
\te
To compute $\Gamma_Q$, we split all fields into a background value plus a fluctuation $X^J\to \bar{X}^J+ x^J$ etc., expand the action eq. (\ref{af}) and discard terms independent or linear in the fluctuations. Then

\be
e^{i\Gamma_Q}=\int Dx\;e^{i\left(S\left[x\right]+\bar S_{bg}\left[\bar X,x\right]+J_{QJ}x^J\right)}
\te
where $S$ is just the action eq. (\ref{af}) evaluated on the fluctuation fields.  $\bar S_{bg}$ is linear on the background fields and quadratic on the fluctuation fields. The sources $J_Q$ enforce the constraints that the expectation value of the fluctuations vanish. For this reason, all one-particle insertions in the diagrammatic evaluation of the effective action cancel out, and it is enough to consider one-particle irreducible graphs only. We shall no longer write the sources explicitly, they are assumed to be included into $S[x]$.

We normalize the integration measure so that

\be
\int Dx\;e^{iS\left[x\right]}=1
\te 
So that $\Gamma_Q[\bar X^J=0]=0$. Then we find 

\be
\frac{\delta \Gamma_Q}{\delta \bar X^J}|_{\bar X^J=0}=\int Dx\;e^{iS\left[x\right]}\frac{\delta S_{bg}}{\delta \bar X^J}=0
\te 
Taking one more derivative

\be 
\frac{\delta^2 \Gamma_Q}{\delta \bar X^J\delta \bar X^K}|_{\bar X^J=0}=i\int Dx\;e^{iS\left[x\right]}\frac{\delta S_{bg}}{\delta \bar X^J}\frac{\delta S_{bg}}{\delta \bar X^K}\equiv i<\frac{\delta S_{bg}}{\delta \bar X^J}\frac{\delta S_{bg}}{\delta \bar X^K}>
\te 

Expanding 

\bea 
S&=&S_0+\tau S_1\nn
S_{bg}&=&S_{bg0}+\tau S_{bg1}
\label{expansion}
\tea
then to first order in $\tau$ we have 

\bea
&&\frac{\delta^2 \Gamma_Q}{\delta \bar X^J\delta \bar X^K}|_{\bar X^J=0}=i<\frac{\delta S_{bg0}}{\delta \bar X^J}\frac{\delta S_{bg0}}{\delta \bar X^K}>_0\nn
&+&i\tau<\frac{\delta S_{bg0}}{\delta \bar X^J}\frac{\delta S_{bg1}}{\delta \bar X^K}>_0+i\tau<\frac{\delta S_{bg1}}{\delta \bar X^J}\frac{\delta S_{bg0}}{\delta \bar X^K}>_0-\tau<\frac{\delta S_{bg0}}{\delta \bar X^J}\frac{\delta S_{bg0}}{\delta \bar X^K}S_1>_0
\label{firstorder}
\tea 
where

\be
\left\langle {X}\right\rangle_0=\int Dx\;e^{iS_0\left[x\right]}{X}
\te
In computing the path integral only one-particle irreducible graphs should be considered. 

The point of this analysis is that the expectation values in eq. (\ref{firstorder}) are computed at $\tau=0$, that is, for a Newtonian theory. We shall not attempt to derive them from the path integral representation of $\Gamma_Q$, but rather assume that they take values that are consistent with K41 theory. We shall come back to discussing the validity of this procedure in the final remarks.

\section{Perturbative energy spectrum}\label{PES}

In this Section we shall derive the energy spectrum to first order in $\tau$, which will be evaluated in next Section \ref{results}. It is convenient to first review the response function and noise kernel in the K41 theory, which will be used later to build the corresponding kernels for viscoelastic flow.

\subsection{Response and noise kernels in K41 theory}

The K41 theory is built on the observation that turbulent fluctuations are non-Gaussian, with skewness

\be 
\left\langle \left[\hat r_j\left(v^j\left(\bm{r}\right)-v^j\left(\bm{0}\right)\right)\right]^3\right\rangle=-\frac 45\epsilon r
\label{ff}
\te
This is the so-called Kolmogorov's $4/5$ law (\cite{FRIS95}). The dynamically generated scale $\epsilon$ has dimensions of $L^2T^{-3}$. It measures the transport of energy accross the turbulent cascade because of nonlinear interactions. 

The K41 theory recognizes three flow regimes. There is a scale $L\approx k_c^{-1}$, basically the linear dimension of the flow, where energy is being injected. This scale and larger wavelengths form the ``creation range''. Out of $\epsilon$ and the molecular viscosity $\nu$ we can form the Kolmogorov scale 

\be 
k_K=(\frac{\epsilon}{\nu^3})^{1/4}
\te 
At smaller length scales the flow is dominated by viscosity. This is the dissipation range. 

From $k_c$ to $k_K$ we are in the inertial range, where $\epsilon$ is the only relevant dimensionful parameter. We may use $\epsilon$ to build the second derivative of the quantum action

\be 
\frac{\delta\Gamma_{Q0}}{\delta\bar A_j(x,t)\delta\bar V^k(y,t')}=\int\frac{d^3k}{(2\pi)^3}\frac{d\omega}{(2\pi)}e^{i[k(x-y)-\omega(t-t')]}\Delta^k_j\zeta\left(\epsilon k^2\right)^{1/3},
\label{sot}
\te
The $0$ subscript denotes that this result holds for $\tau=0$, $\zeta$ is a dimensionless constant, and

\be
\Delta^{j}_k=\delta^j_k-\frac{k^jk_k}{k^2}
\te
is a projector that enforces incompressibility. Adding this to the linearized NSE we obtain the response function \cite{O70,MCCO94}

\bea
\left\langle v^j\left(x,t\right)a_k\left(x',t'\right)\right\rangle_0&=&(-1)\int\frac{d^3k}{(2\pi)^3}\frac{d\omega}{(2\pi)}\frac{e^{i[k(x-y)-\omega(t-t')]}}{[\omega+i\kappa_k]}\Delta^k_j\nn
&=&i\int\frac{d^3k}{(2\pi)^3}\Delta^j_ke^{\left[i\bm{k(x-x')}-\kappa_k\left(t-t'\right)\right]}\theta\left(t-t'\right)
\label{KG}
\tea
where 

\be 
\kappa_k=\zeta\left(\epsilon k^2\right)^{1/3}+\nu k^2
\label{kappa}
\te 
For the noise kernel, we assume 

\be 
\frac{\delta\Gamma_{0}}{\delta\bar A_j(x,t)\delta\bar A_k(y,t')}=i\delta(t-t')\int\frac{d^3k}{(2\pi)^3}e^{i\bm{k(x-y)}}\Delta^{kj}N_k
\label{vot}
\te
and then we may compute the correlation function from eq. (\ref{physcors})

\be
\left\langle v^j\left(x,t\right)v^k\left(x',t'\right)\right\rangle_0=\int\frac{d^3k}{(2\pi)^3}\Delta^j_ke^{\left[i\bm{k(x-x')}-\kappa_k\left|t-t'\right|\right]}\frac{N_k}{2\kappa_k}
\label{KG1}
\te
Taking the coincidence limit yields the energy spectrum

\bea
E_0&=&\int\; dk\;E_0[k]\nn
&=&\frac12<v^2>=\frac1{(2\pi)^2}\int\; dk\;\frac{k^2N_k}{\kappa_k}
\label{KE}
\tea
$A_j$ has dimensions of $L^{-4}T$ and $N_k$ has dimensions of $L^5T^{-3}$. Since we are working in a frame where the fluid is globally at rest, see Appendix (\ref{RGI}), we expect $N_k\to 0$ when $k\to 0$. $N_k$ peaks at the scale $k_c$. In the inertial range, $N_k$ may depend only on $\epsilon$ and $k$, so $N_k\approx \epsilon/k^3$. 

\begin{center}
\begin{figure}[ht]
\scalebox{0.8}{\includegraphics{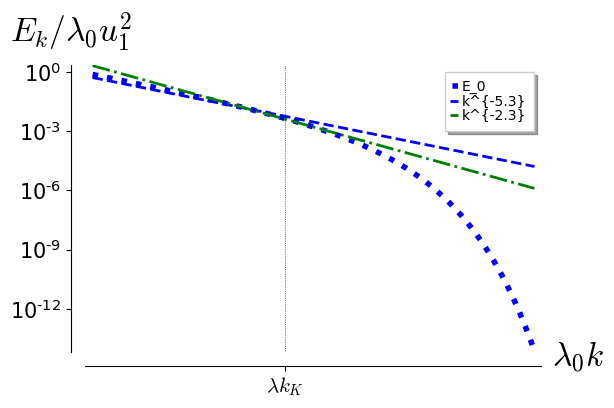}}
\caption{(Color online) A typical K41  energy spectrum as given by eq. (\ref{KS}). For this particular plot we chose ${\rm{Re}}_0=435$. The dotted line is the K41 spectrum; we have included two more lines, dashed corresponding to  $E\approx k^{-5/3}$ and dashed-dotted corresponding $E\approx k^{-2.3}$ for comparison. We see that for moderate Reynolds number, the full spectrum can be fitted to power laws other than $5/3$.}
\label{KSfig}
\end{figure}
\end{center}

$E_0[k]\approx k^4$ in the creation range \cite{Karman48} and falls off exponentially in the dissipation range \cite{Kraichnan59}. We interpolate between these regimes adopting \cite{Hinze75,Pop00}

\be
E_0[k]=\frac{C_K\epsilon^{2/3} k^{4}}{\left( k_c^2+k^2\right)^{17/6}}e^{-\beta(k/k_K)}
\label{KS}
\te 
where $C_K\approx 0.5$ is the Kolmogorov constant  \cite{Sre95}. See fig. (\ref{KSfig}). 

Then

\be 
N_k=\frac{(2\pi)^2C_K\epsilon^{2/3} k^{2}\kappa_k}{\left( k_c^2+k^2\right)^{17/6}}e^{-\beta(k/k_K)}
\label{Nk}
\te 
where $\beta\approx 1/2$ is a dimensionless constant.  Note that the value of $\zeta$ remains undetermined.

In this paper we shall only consider two-point correlations. For the constraints that K41 theory puts on higher correlations see (\cite{PR54,ChM07,KZ18}).

\subsection{Noise kernels in viscoelastic flow}

After reviewing the K41 theory, we return to the viscoelastic flow.

Let us begin by writing in full the expression for the velocity-velocity correlation, eq. (\ref{physcors}). Since the ``classical'' action eq. (\ref{af}) is linear on the auxiliary fields, the noise kernels come entirely from $\Gamma_Q$, whereby

\bea 
&&<v^j(x,t)v^k(x',t')=\int d^3yds\;d^3y'ds'\nn
&&[<v^j(x,t)a_l(y,s)><v^k(x',t')a_m(y',s')>\frac{\delta^2\Gamma_Q}{\partial\bar A_l(y,s)\partial\bar A_m(y',s')}\nn
&+&<v^j(x,t)a_l(y,s)><v^k(x',t')b_{mn}(y',s')>\frac{\delta^2\Gamma_Q}{\partial\bar A_l(y,s)\partial\bar B_{mn}(y',s')}\nn
&+&<v^j(x,t)b_{lp}(y,s)><v^k(x',t')a_m(y',s')>\frac{\delta^2\Gamma_Q}{\partial\bar B_{lp}(y,s)\partial\bar A_m(y',s')}\nn
&+&<v^j(x,t)b_{lp}(y,s)><v^k(x',t')b_{mn}(y',s')>\frac{\delta^2\Gamma_Q}{\partial\bar B_{lp}(y,s)\partial\bar B_{mn}(y',s')}]
\label{fullcors}
\tea
Our first concern will be to show that, of the four noise kernels listed in eq. (\ref{fullcors}), only the first survives in the high Reynolds number limit, where viscosity effects are negligible. In other words, when $\nu\to 0$

\be
\frac{\delta^2\Gamma_Q}{\partial\bar A_l(y,s)\partial\bar B_{mn}(y',s')}\approx \frac{\delta^2\Gamma_Q}{\partial\bar B_{lp}(y,s)\partial\bar B_{mn}(y',s')}\approx 0
\label{trivialnoises}
\te
Let us begin by listing $S_0$, $S_1$, $S_{bg0}$ and $S_{bg1}$, cfr. eqs (\ref{expansion}), in the $\nu=0$ limit.

\bea
S_0&=&\int d^3yds\;\left[a_j\left(v^j_{,t}+v^kv^j_{,k}+p^{jk}_{,k}\right)+b_{jk}p^{jk}\right]\nn
S_1&=&\int d^3yds\;b_{jk}\left(p^{jk}_{,t}+v^lp^{jk}_{,l}-p^{jl}v^k_{,l}-v^j_{,l}p^{lk}\right)
\tea 

\bea 
S_{bg0}&=&\int d^3yds\;\left[\bar A_jv^kv^j_{,k}+\bar V^j(a_kv^k_{,j}-a_{j,k}v^k)\right]\nn
S_{bg1}&=&\int d^3yds\;[\bar B_{jk}\left(v^lp^{jk}_{,l}-p^{jl}v^k_{,l}-v^j_{,l}p^{lk}\right)\nn
&+&\bar V^j(b_{kl}p^{kl}_{,j}+2\partial_l(b_{jk}p^{kl})-\bar P^{jk}(v^lb_{jk,l}+2b_{jl}v^l_{,k})]
\tea 
Let us return to verifying eq. (\ref{trivialnoises}). Since $\delta S_{bg0}/\delta\bar B_{jk}=0$, there is simply nothing to compute when applying eq. (\ref{firstorder}) to $\delta^2\Gamma_Q/\delta\bar B_{jk}\delta\bar B_{lm}$, which must be at least 0f $O(\tau^2)$. 

With respect to $\delta^2\Gamma_Q/\delta\bar A_{j}\delta\bar B_{kl}$, from eq. (\ref{firstorder})  we find 

\bea 
&&\frac{\delta^2\Gamma_Q}{\delta\bar A_j(y,s)\delta\bar B_{kl}(y',s')}=i\tau<\frac{\delta S_{bg0}}{\delta\bar A_j(y,s)}\frac{\delta S_{bg1}}{\delta\bar B_{kl}(y',s')}>_0\nn
&=&<(v^mv^j_{,m})(y,s)\left(v^np^{kl}_{,n}-p^{kn}v^l_{,n}-v^k_{,n}p^{nl}\right)(y',s')>_0
\label{toomanyps}
\tea
It is not necessary to filter out the longitudinal modes, since the whole expression vanishes. Because $S_0$ only contains $b_{jk}$ in the combination $b_{jk}p^{jk}$, we find a Novikov-type formula \cite{Novikov}.

\be
\left\langle p_{jk}{X}\right\rangle_0=i\left\langle \frac{\delta{X}}{\delta b_{jk}}\right\rangle_0
\label{Nov}
\te
From this formula it is obvious that eq. (\ref{toomanyps}) vanishes. 

It is an important point that we have been able to show the validity of eq. (\ref{trivialnoises}) directly from the properties of the path integral, without relying on any particular turbulent flow model. However, in the remaining, we shall need to rely on the K41 theory to move forward.

Since only the first line of eq. (\ref{fullcors}) survives, we only need one response function, namely $<v^j(x,t)a_l(y,s)>$, and one noise kernel, namely $\delta^2\Gamma_Q/\delta\bar A_j\delta\bar A_k$. We shall consider the latter here, and the former in next section.

At $\tau=0$ the noise kernel is given by eqs. (\ref{vot}) and (\ref{Nk}). We shall now show that the linear order corrections from eq. (\ref{firstorder}) vanish. Indeed since $\delta S_{bg1}/\delta \bar A_j=0$, the only possible correction is

\be 
<\frac{\delta S_{bg0}}{\delta \bar A_j(x,t)}\frac{\delta S_{bg0}}{\delta \bar A_k(x',t'}S_1>_0=\int d^3yds<(v^lv^j_{,l})(x,t)(v^mv^k_{,m})(x',t')(b_{np}\left(p^{np}_{,t}+v^qp^{np}_{,q}-p^{nq}v^p_{,q}-v^p_{,q}p^{qn}\right))(y,s)>
\te 
which is seen to vanish from eq. (\ref{Nov}). Once again, we do not filter our the longitudinal modes.

\subsection{The response functions}

The remaining step to compute the viscoelastic spectrum is to find the response function $<v^j(x,t)a_k(x',t')>$. To do this we must solve the system 

\bea
&&\int d^3ydt'\left\{\frac{\delta^2\Gamma}{\delta \bar A_j\left(x,t\right)\delta \bar V^k\left(y,t'\right)}\left\langle v^k\left(y,t'\right)a_l\left(x',t''\right)\right\rangle
+\frac{\delta^2\Gamma}{\delta \bar A_j\left(x,t\right)\delta \bar P^{km}\left(y,t'\right)}\left\langle p^{km}\left(y,t'\right)a_l\left(x',t''\right)\right\rangle\right\}=i\Delta^j_l\nn
&&\int d^3ydt'\left\{\frac{\delta^2\Gamma}{\delta \bar B_{jn}\left(x,t\right)\delta \bar V^k\left(y,t'\right)}\left\langle v^k\left(y,t'\right)a_l\left(x',t''\right)\right\rangle
+\frac{\delta^2\Gamma}{\delta \bar B_{jn}\left(x,t\right)\delta \bar P^{km}\left(y,t'\right)}\left\langle p^{km}\left(y,t'\right)a_l\left(x',t''\right)\right\rangle\right\}=0
\label{third}
\tea
It is easy to see that the first order corrections to $\delta^2\Gamma /\delta \bar A_j\left(x,t\right)\delta \bar V^k\left(y,t'\right)$, $\delta^2\Gamma /\delta \bar B_{jn}\left(x,t\right)\delta \bar V^k\left(y,t'\right)$ and $\delta^2\Gamma /\delta \bar B_{jn}\left(x,t\right)\delta \bar P^{km}\left(y,t'\right)$ vanish because of eq. (\ref{Nov}). 

The only kernel we need to compute the Schwinger-Dyson equations to first order in $\tau$ is 

\be
\frac{\delta^2\Gamma_Q}{\delta \bar A_j\left(x,t\right)\delta \bar P^{km}\left(y,t'\right)}=-i\tau\left\langle \left(v^lv^j_{,l}\right)\left(x,t\right)\left[v^nb_{km,n}+b_{kn}v^n_{,m}+b_{mn}v^n_{,k}\right]\left(y,t'\right)\right\rangle_0
\te
Where only transverse modes contribute to the variational derivative with respect to $\bar A_j\left(x,t\right)$.
We assume this kernel is local in time, and then on dimensional grounds ($P^{km}$ has units of $L^2T^{-2}$, $B_{km}$ has units of $L^{-5}T$)

\be
\frac{\delta^2\Gamma_Q}{\delta \bar A_j\left(x,t\right)\delta \bar P^{km}\left(y,t'\right)}=-\frac i2\tau\delta(t-t')\int\frac{d^3k}{(2\pi)^3}e^{i\bm{k(x-y)}}\left[\Delta^k_jk^m+\Delta^m_jk^k\right]\Lambda\kappa_k
\label{uot}
\te
where $\Lambda$ is dimensionless 

\be
\Lambda\approx \frac {2C_K}{3}\frac{k^2\epsilon^{2/3}}{\kappa_k^2k_c^{2/3}}
\label{Lambda1}
\te
See appendix (\ref{Lambdacomp}).

We may now solve the system eq. (\ref{third}). Let us parametrize

\bea
&&\left\langle v^j\left(x,t\right)a_k\left(x',t'\right)\right\rangle=\int\frac{d^3k}{(2\pi)^3}\frac{d\omega}{(2\pi)}e^{i\left[\bm{k(x-x')}-\omega\left(t-t'\right)\right]}\Delta^j_kG\left[k,\omega\right]\nn
&&\left\langle p^{jm}\left(x,t\right)a_k\left(x',t'\right)\right\rangle=i\int\frac{d^3k}{(2\pi)^3}\frac{d\omega}{(2\pi)}e^{i\left[\bm{k(x-x')}-\omega\left(t-t'\right)\right]}\left[\Delta^j_kk^m+\Delta^m_kk^j\right]G'\left[k,\omega\right]
\tea 

Then

\bea
\left[-i\omega+\zeta(\epsilon k^2)^{1/3}\right]G-k^2\left[1-\Lambda\kappa_k\tau\right]G'&=&i\nn
\nu G+\left[-i\omega\tau+1\right]G'&=&0
\tea
Solving for  $G$

\be
G[k,\omega]=(-i)\frac{[1-i\omega\tau]}{P[\omega]}
\te
where 

\be 
P[\omega]=\tau\omega^2+i\omega[1+\tau\zeta(\epsilon k^2)^{1/3}]-\kappa_k[1-\Lambda\tau\nu k^2]
\label{Pomega}
\te
Once $G$ is known, we may compute

\be
\left\langle v^j\left(x,t\right)v^k\left(x',t'\right)\right\rangle=\int\frac{d^3k}{(2\pi)^3}\frac{d\omega}{(2\pi)}e^{i\left[\bm{k(x-x')}-\omega\left(t-t'\right)\right]}\Delta^{jk}G_1\left[k,\omega\right]
\te
where

\be
G_1\left[k,\omega\right]=\left|G\left[k,\omega\right]\right|^2N_k
\te
$N_k$ as in eq. (\ref{Nk}). The energy spectrum is computed from the coincidence limit of the velocity self-correlation

\be
E\left[k\right]=\frac1{\left(2\pi\right)^2} k^2N_k\int \frac{d\omega}{(2\pi)}\left|G\left[k,\omega\right]\right|^2\equiv E_0[k]\frac{[1+\kappa_k\tau+\Lambda\tau\nu k^2]}{\left[1+\zeta\tau(\epsilon k^2)^{1/3}\right]}
\label{spectrum}
\te 
$E_0$ is defined in eq. (\ref{KS}), $\Lambda$ in eq. (\ref{Lambda1}). See the details in Appendix \ref{spectcomp}
\section{Results}\label{results}

To be able to compare the spectrum found in eq. (\ref{spectrum}) with results in the literature we need to introduce several relevant scales.

The spectrum eq. (\ref{spectrum}) defines a velocity scale 

\be 
u_1^2=\frac23\int\;dk\;E\left[k\right]\approx \frac23C_K(\frac{\epsilon}{k_c})^{2/3}
\label{u1}
\te 
We also introduce the Taylor microscale $\lambda_0$ \cite{TL}

\be 
\frac1{\lambda_0^2}=\frac{10}{u_1^2}\int\;dk\;k^2\;E\left[k\right]\approx 10k_c^{2/3}k_K^{4/3}
\label{lambda0}
\te 
and two dimensionless parameters, the Newtonian Reynolds number

\be
{\rm{Re}}=\frac{u_1\lambda_0}{\nu}\approx\sqrt{\frac{C_K}{15}}(\frac{k_K}{k_c})^{2/3}
\label{Re}
\te
and the Weissenberg number \cite{MS15} 

\be
{\rm{Wi}}=\frac{\tau\nu}{\lambda^2_0}
\label{Wi}
\te
From eqs. (\ref{lambda0}) and (\ref{Re}) we obtain 

\bea
&&\lambda_0k_c\approx\sqrt{\frac{C_K}{150}}\frac1{{\rm{Re}}}\nn
&&\lambda_0k_K\approx(\frac3{20C_K})^{1/4}{\rm{Re}}^{1/2}
\tea
We may now write

\bea
\nu k^2\tau&=&{\rm{Wi}}\;(\lambda_0k)^2\nn
(\epsilon k^2)^(1/3)&=&(\frac 3{20C_K})^{1/3}{\rm{Re}}^{2/3}{\rm{Wi}}(\lambda_0k)^{2/3}
\tea 
and 

\be
\Lambda\approx \frac{32}9C_K^{2/3}\frac{{\rm{Re}}^{2/3}(\lambda_0k)^{2/3}}{[\zeta+(\frac{20C_K}{3{\rm{Re}}^2})^{1/3}(\lambda_0k)^{4/3}]^2}
\label{Lambdafin}
\te

Observe that in the terms of the viscoelastic model presented in ref. \cite{ACR23} we are working in the limit where the Bingham number goes to infinity.

\begin{center}
\begin{figure}[ht]
\scalebox{0.8}{\includegraphics{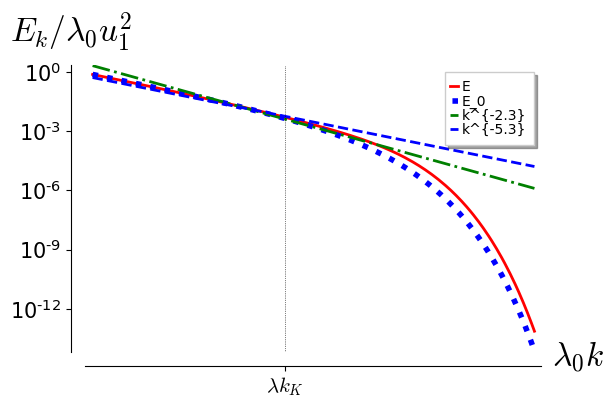}}
\caption{(Color online) A typical energy spectrum as given by eq. (\ref{spectrum}). For this particular plot we chose ${\rm{Re}}_0=435$, ${\rm{Wi}}=10^{-3}$ and the constant $\zeta$ in eq. (\ref{kappa}) as $\zeta=17.5$. The full line is the actual spectrum; we have included three more lines, dotted corresponding to the Kolmogorov spectrum eq. (\ref{KS}), dashed corresponding to $E\approx k^{-5/3}$ and dashed-dotted corresponding to $E\approx k^{-2.3}$ for comparison. These last three are the same as in fig. (\ref{KSfig}). We have normalized the spectrum so that it converges to the universal Kolmogorov spectrum for $k\approx \lambda_0^{-1}$. }
\label{fig1}
\end{figure}
\end{center}

\begin{center}
\begin{figure}[ht]
\scalebox{0.8}{\includegraphics{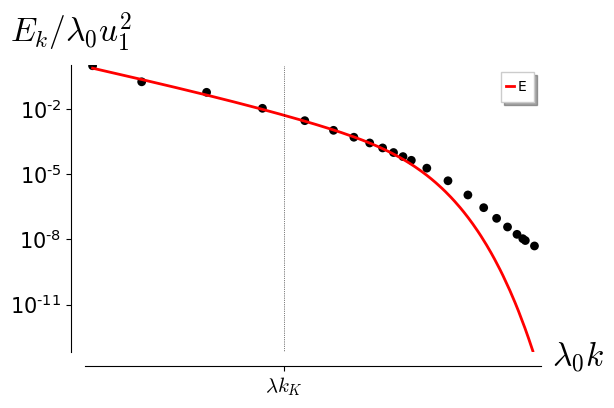}}
\caption{(Color online) We plot the same non-Newtonian spectrum as in fig. (\ref{fig1}), with ${\rm{Re}}_0=435$, ${\rm{Wi}}=10^{-3}$ and $\zeta=17.5$ (full line). Dots are data from the numerical simulations reported in  fig. 2 of \cite{ACR23}, corresponding to the largest Bingham number available. Data read from \cite{Rostidata}}.
\label{fig3}
\end{figure}
\end{center}

We plot a typical spectrum in fig. (\ref{fig1}), and then compared it to simulation data extracted from ref. \cite{ACR23} in fig (\ref{fig3}).

\section{Final remarks}\label{Final}

Our goal in this paper is to develop a practical tool to explore problems that can be regarded as perturbations of homogeneous, isotropic turbulence in Newtonian incompressible fluids. 

To achieve this goal we have deployed functional methods, such as the EA approach to the MSR formalism, to obtain formal expressions that display the parametric dependence of the correlation functions in full. These formal expressions can be used to construct perturbative series in terms of a suitable small parameter, the Weissenberg number in the case of viscoelastic flow, or the inverse speed of light for a relativistic fluid. 

Of course, to obtain results which may be matched against numerical and experimental data requires computing the expansion coefficients to a high accuracy. Achieving this accuracy  may be regarded as an unsolved problem in itself. We sidestep this issue by relying on our understanding of fully developed turbulence. Thus we replace the formal path integrals by simple ansätze known to work in Kolmogorov turbulence. We regard this procedure as an extreme, but valid, kind of renormalization \cite{MCCO04}. 

As a final remark on this subject, we note that the application of field theory methods to turbulence is almost as old as the K41 theory itself \cite{EF10,Wyld61,Lee65,LB87,LP96,BLP98,LP98,V98,BSMCC13,AK24}. The MSR approach is typically seen as the most systematic way to translate  a classical stochastic problem into a field theory. It has been used extensively in turbulence (to cite a few examples, see \cite{BL87,E94,T97,Tom97,FF24}). However, there remain lingering doubts on whether these methods can provide a convincing derivation of the K41 theory, let alone a means to improve it, such as providing a better prediction of the scaling exponents for the higher structure functions \cite{K71,GBF81,G98,Tsin09,MC23,MC24}. Replacing the unknown correlations by their K41 form is a way to mimic the yet undiscovered non-perturbative methods necessary to compute them from first principles.

We would like to conclude with a few comments on our results. 

It is customary to fit turbulent spectra to power laws and in this sense a $7/3$ exponent for velocity correlations in Non Newtonian fluids has been reported, see eg. \cite{MT21,ACR23}. This scaling does not seem to emerge from our analysis. However, as fig. (\ref{KSfig}) shows, for moderate Reynolds numbers the spectrum spans a range of scaling laws in the transition from the inertial to the dissipative range. For ${\mathrm{Re}}=435$ (as shown in fig. (\ref{KSfig})) it actually displays a $7/3$ scaling around the Kolmogorov scale. The Non-Newtonian spectrum in fig. (\ref{fig1}) shows a similar behavior.

This could be a case where two completely different analytic expressions nevertheless yield similar results. This is not uncommon, the agreement between the Colebrook formula and the Blasius law for the friction factor in pipe turbulence comes to mind \cite{SCH79}. 

A remarkable property of the spectrum eq. (\ref{spectrum}), clearly shown in fig. (\ref{fig1}), is the enhancement of velocity fluctuations for very small scales, deep in the dissipative range, relative to the already strongly suppressed Newtonian spectrum of eq. (\ref{KS}). To the best of our understanding, this is not ruled out by available data, such as those in refs. (\cite{Zhang21}), (\cite{MT21}) and (\cite{ACR23}), but it may well be a make or break issue for the validation of the model. 

Regarding this last point, we believe that the most relevant feature of our results is the way they display the functional dependence of the non-Newtonian spectrum on the different parameters of the theory. Validating this theoretical prediction requires comparing it against numerical and experimental data accross a full range of Reynolds and Weissenberg numbers.

We stress that we have discussed flows which are strictly homogeneous and isotropic. Data obtained from different configurations, such as wall-bounded flow (\cite{MT21}), while relevant, cannot be directly compared to our theory. 

In conclusion, our goal is to establish the cogency of our proposal, with its validation to be discussed in future communications.

\acknowledgements

I thank P. Mininni for multiple talks.

E. C. acknowledges financial support from Universidad de Buenos Aires through Grant No. UBACYT
20020170100129BA, CONICET Grant No. PIP2017/19:11220170100817CO and ANPCyT Grant No. PICT 2018: 03684. 

\appendix 

\section{Non-Newtonian fluid as the non-relativistic limit of a conformal fluid}\label{CM}

We consider a relativistic fluid of massless particles. 

At the macroscopic level, the theory is described by the energy-momentum tensor (EMT) $T^{\mu\nu}$. Adopting the Landau prescription for the four velocity $u^{\mu}$ and the energy density $\rho$ 

\be
T^{\mu}_{\nu}u^{\nu}=-\rho u^{\mu}
\te
and observing that $T^{\mu\nu}$ is traceless, we are led to write

\be
T^{\mu\nu}=\rho\left[u^{\mu}u^{\nu}+\frac13\Delta^{\mu\nu}+\Pi^{\mu\nu}\right]
\label{tmunu}
\te
where

\be
\Delta^{\mu\nu}=\eta^{\mu\nu}+u^{\mu}u^{\nu}
\te 
and

\be
\Pi^{\mu}_{\nu}u^{\nu}=\Pi^{\mu}_{\mu}=0
\label{TT}
\te
We must also provide an entropy flux. For an ideal fluid, namely when $\Pi^{\mu}_{\nu}=0$, the entropy density is 

\be
s=s_0=\frac1T\left(\rho+P\right)
\te
where $T$ is the temperature and $P=\rho/3$ is the pressure. From the thermodynamic relation

\be
s_0=\frac{\partial P}{\partial T}
\te 
we conclude that

\be
\rho=\sigma_{SB} T^4
\te
for some constant $\sigma_{SB}$. The entropy flux is then

\be
S_0^{\mu}=s_0u^{\mu}
\te
When we consider the real fluid, $\Pi^{\mu}_{\nu}\not=0$, we observe that because of (\ref{TT}) we cannot make a vector out of $u^{\mu}$ and $\Pi^{\mu\nu}$. Therefore it makes sense to write

\be
S^{\mu}=su^{\mu}
\te
The entropy density ought to be maximum when the fluid is in equilibrium, namely when $\Pi^{\mu\nu}$ vanishes. So at least close to equilibrium we should have

\be
s=\frac43\sigma_{SB}T^3e^{-\frac32\lambda\Pi^{\mu\nu}\Pi_{\mu\nu}}
\te 
for some dimensionless constant $\lambda$. If we further write 

\be
T=T_0e^{\delta}
\te
Then the conservation laws are 

\bea
0&=&\delta_{,\nu}u^{\nu}+\frac13u^{\nu}_{,\nu}+\frac14\Pi^{\mu\nu}u_{\mu,\nu}\nn
0&=&\delta_{,\nu}\left[\Delta^{\mu\nu}+3\Pi^{\mu\nu}\right]+u^{\mu}_{,\nu}u^{\nu}+\frac34\Delta^{\mu}_{\rho}\Pi^{\rho\nu}_{,\nu}
\label{cons}
\tea 
On the other part, positive entropy creation yields

\be
0\le\frac13S^{\mu}_{\mu}=u^{\nu}\left[\delta_{,\nu}-\lambda\Pi^{\rho\sigma}\Pi_{\rho\sigma,\nu}\right]+\frac13u^{\nu}_{,\nu}
\te
which using the conservation laws and the transversality of $\Pi^{\rho\sigma}$ may be written as

\be
\Pi^{\rho\sigma}\left[\lambda u^{\nu}\Pi_{\rho\sigma,\nu}+\frac18\sigma_{\rho\sigma}\right]\le 0
\te
where

\be
\sigma^{\rho\sigma}=\left[\Delta^{\rho\mu}\Delta^{\sigma\nu}+\Delta^{\rho\nu}\Delta^{\sigma\mu}-\frac23\Delta^{\rho\sigma}\Delta^{\mu\nu}\right]u_{\mu,\nu}
\te
is the covariant form of the shear tensor eq. (\ref{shear}).
Therefore, positive entropy creation is achieved by adopting the Cattaneo-Maxwell equation

\be
\lambda u^{\nu}\Pi^{\rho\sigma}_{,\nu}+\frac1{t_R}\Pi^{\rho\sigma}+\frac18\sigma^{\rho\sigma}=0
\label{CM2}
\te
We shall now consider the nonrelativistic limit. We write explicitly $x^0=ct$ and
\bea
u^{\mu}&=&\frac{\left( 1,u^k/c\right) }{\sqrt{1-u^2/c^2}}\nn
\Pi^{\mu\nu}&=&\left( \begin{array}{cc}\Pi_{lm}u^lu^m/c^2&\Pi_{kl}u^l/c\\\Pi_{jm}u^m/c&\Pi_{jk}\end{array}\right) +\frac{\Pi_{lm}u^lu^m/c^2}{3-u^2/c^2}\left( \begin{array}{cc}u^2/c^2&u^k/c\\u^j/c&\delta_{jk}\end{array}\right)
\tea 
where $\Pi^j_j=0$.  Observe that

\be
\Delta^{\mu}_{\nu}=\frac1{1-u^2/c^2}\left( \begin{array}{cc}-u^2/c^2&u^k/c\\-v_j/c&\delta^{jk}+\left(u^ju^k-u^2\delta^{jk}\right)/c^2\end{array}\right)
\te
The first nontrivial terms in the energy conservation equation are of order $1/c$ and read

\be
0=\delta_{,t}+u^j\delta_{,j}+\frac13u^{j}_{,j}+\frac14\Pi^{jk}v_{j,k}
\te 
From the momentum conservation equation we get

\be
0=\delta_{,k}\left[\delta^{jk}+3\Pi^{jk}\right]+\frac34\Pi^{jk}_{,k}+\frac1{c^2}\left[u^{j}_{,t}+u^{k}u^j_{,k}\right]
\label{ns}
\te
The Cattaneo-Maxwell equation (\ref{CM2}) yields

\be
{\lambda} \left[\Pi^{jk}_{,t}+u^l\Pi^{jk}_{,l}\right]+\frac1{t_R}\Pi^{jk}+\frac1{8}\left(v_{j,k}+v_{k,j}\right)=0
\label{CMNR}
\te
A consistent nonrelativistic limit requires $\delta,\Pi^{jk}\propto 1/c^2$. Then from energy conservation we get $u^j_{,j}=0$ to lowest order. Let us write

\bea 
u_j&=&v_j+\frac1{c^2}\phi_{,j}\nn
\Pi^{jk}&=&\frac4{3c^2}p^{jk}\nn
\delta&=&\frac1{c^2}\epsilon\nn
\lambda&=&\frac{3c^2}{32\nu}\tau\nn
t_R&=&\frac{32\nu}{3c^2}
\tea
where $v^j_{,j}=0$.

Collecting again the leading terms we get 

\be
0=\epsilon_{,t}+v^j\epsilon_{,j}+\frac13\bm{\Delta}\phi+\frac13v^{j,k}p_{jk}
\te 

\be
0=\epsilon_{,j}+p^{jk}_{,k}+\left[v^{j}_{,t}+v^{k}v^j_{,k}\right]
\label{nsnr}
\te
Taking the divergence of this equation we get

\be
0=\bm{\Delta}\epsilon+v^{k,j}v^j_{,k}+p^{jk}_{,jk}
\te
so we may write a scalar-free equation of motion

\be
Q^l=\Delta^{l}_j\left[v^{j}_{,t}+v^{k}v^j_{,k}+p^{jk}_{,k}\right]=0
\te
where

\be
\Delta_{jk}=\delta_{jk}-\partial_j\bm{\Delta}^{-1}\partial_k
\te
Finally

\be
Q^{jk}=\tau \left[p^{jk}_{,t}+v^lp^{jk}_{,l}\right]+p^{jk}+\nu \left[v_{j,k}+v_{k,j}\right]=0
\label{CMNR2}
\te
We may define a mass density

\be
\mu=\frac{\rho}{c^2}
\te 
Then $\mu$ is constant to order $1/c^2$. We see that $\epsilon=P/\mu$, where $P$ is the non-constant part of the pressure. $P$ is not a dynamical variable but it is determined from the constraint

\be
0=\frac1{\mu}{\bm{\Delta}} P+ p^{jk}_{,jk}+v^{k}_{,j}v^j_{,k}
\label{constraint}
\te
We see that we recover equations (\ref{eoms}) in the particular case $\tau_2=0$. 

\section{Random Galilean invariance}\label{RGI}

Let us go back to the action functional eq. (\ref{af}) and the corresponding generating functional eq. (\ref{gf}), whose Legendre transform yields the 1PI effective action $\Gamma$, eq. (\ref{1PIEA}). 

This construction misses the fact that the equations of motion (\ref{eoms}) are \emph{random galilean invariant}, that is, they are invariant under the transformation

\bea
&&v^j\left(x^j,t\right)\to v^j\left(x^j-\epsilon^j\left(t\right),t\right)+\dot\epsilon^j\left(t\right)\nn
&& p^{jk}\left(x^j,t\right)\to p^{jk}\left(x^j-\epsilon^j\left(t\right),t\right)\nn
&&A_j\left(x^j,t\right)\to A_j\left(x^j-\epsilon^j\left(t\right),t\right)\nn
&&A_{jk}\left(x^j,t\right)\to A_{jk}\left(x^j-\epsilon^j\left(t\right),t\right)
\tea
where $\epsilon^j\left(t\right)$ is an arbitrary time dependent field. Of course we are using that

\be 
\int d^3x\;A_j\ddot\epsilon^j=\int d^3x\;A_j\partial^j\left(\ddot\epsilon_kx^k\right)=0
\te
For this reason the path integral defining the generating functional, eq. (\ref{gf}), is redundant. To eliminate the overcounting, we consider the non-invariant function

\be
P^j\left(t\right)=\int d^3x\;\mu v^j
\te 
Assuming that $\mu$ transforms as $\mu\left(x^j,t\right)\to \mu\left(x^j-\epsilon^j\left(t\right),t\right)$ we see that

\be
P^j\to P^j+M\dot\epsilon^j\left(t\right)
\te
where $M$ is the total mass of the fluid. We now observe that

\be 
1=\int\;D\epsilon^j\;{\rm{det}}\frac{\delta P^j\left[\epsilon\right]}{\delta\epsilon^k}\delta\left(P^j\left[\epsilon\right]-C^j\right)
\te 
Introducing this identity into the path integral, we can take the $\epsilon$ integral out as a constant factor (for this we make a change of variables within the integral, with unit Jacobian), integrate over the $C^j$ with a Gaussian weight and exponentiate the determinant introducing Grassmann variables $\zeta_j$ and $\eta^j$, where now 

\be
e^{iW\left[Z_a,H^a,z_a,h^a\right]}=\int\;DX^aDA_a\;e^{i\left(S_{RGI}+Z_aX^a+H^aA_a+z_a\eta^a+h^a\zeta_a\right)}
\label{MSRRGI}
\te 
where

\be 
S_{RGI}=\int dtd^3x\;\left\{A_jQ^j+B_{jk}Q^{jk}\right\}+\frac1{2\alpha}\int\;dt\;P_jP^j+i\int\;dt\;\zeta_jM\dot\eta^j
\te 
Note that the ghost fields are decoupled. This action is still invariant under a BRST transformation defined as follows: the matter and auxiliary fields transform as in a random galilean transformation with parameter $\epsilon^j=\theta\eta^j$, where $\theta$ is a Grassmann constant, $\zeta_j$ transforms into $\zeta_j+i\theta P_j/\alpha$, and $\eta^j$ is invariant. We thus obtain the Zinn-Justin equation \cite{Zin93}

\bea 
&&\int\;d^dxdt\;\left\{\frac{\delta\Gamma}{\delta \bar V^j}\left(\left\langle \eta^l\left(t\right)v^j_{,l}\left(x^l,t\right)\right\rangle-\dot{\bar\eta}^j\left(t\right)\right)+\frac{\delta\Gamma}{\delta \bar P^{jk}}\left\langle \eta^l\left(t\right) p^{jk}_{,l}\left(x^l,t\right)\right\rangle+\frac{\delta\Gamma}{\delta \bar A_j}\left\langle \eta^l\left(t\right)a_{j,l}\left(x^l,t\right)\right\rangle\right.\nn
&+&\left.\frac{\delta\Gamma}{\delta \bar B_{jk}}\left\langle \eta^l\left(t\right)b_{jk,l}\left(x^l,t\right)\right\rangle\right\}-\frac i{\alpha}\int\;dt\;\frac{\delta\Gamma}{\delta \bar\zeta_j}\left\langle P_j\left(t\right)\right\rangle=0
\label{ZJE}
\tea
Since the integral over ghost fields is just a decoupled Gaussian integral, the binary products decouple, namely

\be
\left\langle \eta^l\left(t\right)v^j_{,l}\left(x^l,t\right)\right\rangle=\bar\eta^l\left(t\right)\bar V^j_{,l}\left(x^l,t\right)
\te
etc., and

\be
\frac{\delta\Gamma}{\delta\bar \zeta_j}=iM\dot{\bar\eta}^j
\te
Moreover

\be 
\int\;d^dxdt\;\bar\eta^l\left(t\right)\bar V^j_{,l}\left(x^l,t\right)\frac{\delta}{\delta v^j}\int\;dt\;P_k\left(t\right)P^k\left(t\right)=0
\te
So eq. (\ref{ZJE}) is consistent with 

\be
\Gamma=\Gamma_0+\frac1{2\alpha}\int\;dt\;P_j\left(t\right)P^j\left(t\right)
\te 
where $\Gamma_0$ is independent of $\alpha$. $\Gamma_0$ is an effective action without the Fadeev-Popov procedure. This implies that $\Gamma_0$ is identically zero when the auxiliary fields vanish, independently of the physical fields.

In the presence of the gauge-fixing term  $\left\langle vA\right\rangle$ and $\left\langle vv\right\rangle$ are unchanged, and now

\be
\left\langle AA\right\rangle=\frac{\mu^2}{\alpha}\frac{\delta\left(k\right)}{\omega^2+\nu^2\left[0\right]}
\te
When $\alpha\to 0$ this forces the noise kernels and the self energies to vanish at zero momentum, as we have assumed in the text.

\section{Derivation of eq. (\ref{Lambda1})}\label{Lambdacomp}

We may estimate $\Lambda$ as follows. First note that at $\tau=0$ $p^{jk}$ becomes a Lagrange multiplier enforcing the constraint

\be
b_{jk}=\frac12\left[a_{j,k}+a_{k,j}\right]
\te
Now use a quasi-Gaussian approximation to get

\bea
&&\frac{\delta^2\Gamma_Q}{\delta \bar A_j\left(x,t\right)\delta \bar P^{km}\left(y,t'\right)}=-\frac{i\tau}2\left\langle \left(v^lv^j_{,l}\right)\left(x,t\right)\left[v^n\left[a_{k,mn}+a_{m,kn}\right]+\left[a_{k,n}+a_{n,k}\right]v^n_{,m}+\left[a_{m,n}+a_{n,m}\right]v^n_{,k}\right]\left(y,t'\right)\right\rangle_0\nn
&=&-\frac{i\tau}2\left\langle \left(v^lv^j_{,l}\right)\left(x,t\right)\left[\partial^2_{mn}\left(v^na_{k}\right)+\partial^2_{kn}\left(v^na_{m}\right)+a_{n,k}v^n_{,m}+a_{n,m}v^n_{,k}\right]\left(y,t'\right)\right\rangle_0\nn
&=&-\frac{i\tau}2\left\{\frac{\partial^2}{\partial y^m\partial y^n}\left[\left\langle v^l\left(x,t\right)v^n\left(y,t'\right)\right\rangle\left\langle v^j_{,l}\left(x,t\right)a_k\left(y,t'\right)\right\rangle+\left\langle v^l\left(x,t\right)a_k\left(y,t'\right)\right\rangle\left\langle v^j_{,l}\left(x,t\right)v^n\left(y,t'\right)\right\rangle\right]\right.\nn
&+&\left.\left\langle v^l\left(x,t\right)a_{n,k}\left(y,t'\right)\right\rangle\left\langle v^j_{,l}\left(x,t\right)v^n_{,m}\left(y,t'\right)\right\rangle+\left\langle v^l\left(x,t\right)v^n_{,m}\left(y,t'\right)\right\rangle\left\langle v^j_{,l}\left(x,t\right)a_{n,k}\left(y,t'\right)\right\rangle+\left(k\leftrightarrow m\right)\right\}
\tea
We replace eqs. (\ref{KG}) and (\ref{KG1})

\bea
&&\frac{\delta^2\Gamma_Q}{\delta \bar A_j\left(x,t\right)\delta \bar P^{km}\left(y,t'\right)}=-i\frac{\tau}2\theta\left(t-t'\right)\int\frac{d^3k}{(2\pi)^3}e^{i\bm{k(x-y)}}\nn
&&\left[ \left(k_mk_n\right)\int\frac{d^3k'}{(2\pi)^3}\Delta_{k'}^{ln}e^{-\left[\kappa_{k'}+\kappa_{\left(k-k'\right)}\right]\left(t-t'\right)}\left(k-k'\right)_l\Delta^j_{\left(k-k'\right)k}\frac{N_{k'}}{2\kappa_{k'}}\right.\nn
&+&\left(k_mk_n\right)\int\frac{d^3k'}{(2\pi)^3}\Delta_{k'}^{jn}e^{-\left[\kappa_{k'}+\kappa_{\left(k-k'\right)}\right]\left(t-t'\right)}k'_l\Delta^l_{\left(k-k'\right)k}\frac{N_{k'}}{2\kappa_{k'}}\nn
&+&\int\frac{d^3k'}{(2\pi)^3}\Delta_{k'}^{jn}e^{-\left[\kappa_{k'}+\kappa_{\left(k-k'\right)}\right]\left(t-t'\right)}k'_lk'_m\Delta^l_{\left(k-k'\right)n}\left(k-k'\right)_k\frac{N_{k'}}{2\kappa_{k'}}\nn
&+&\left.\int\frac{d^3k'}{(2\pi)^3}\Delta_{k'}^{ln}e^{-\left[\kappa_{k'}+\kappa_{\left(k-k'\right)}\right]\left(t-t'\right)}k'_m\Delta^j_{\left(k-k'\right)n}\left(k-k'\right)_l\left(k-k'\right)_k\frac{N_{k'}}{2\kappa_{k'}}\right]+\left(k\leftrightarrow m\right)
\tea
Since the $k'$ integral is dominated by the infrared band, we approximate $k'\ll k$. Observe that two of the integrals vanish from symmetry, so

\bea
&&\frac{\delta^2\Gamma_Q}{\delta \bar A_j\left(x,t\right)\delta \bar P^{km}\left(y,t'\right)}=-i\frac{\tau}2\theta\left(t-t'\right)\int\frac{d^3k}{(2\pi)^3}e^{i\bm{k(x-y)}}e^{-\kappa_{k}\left(t-t'\right)}\nn
&&\left[ \left(k_mk_n\right)k_l\Delta^j_{\left(k\right)k}\int\frac{d^3k'}{(2\pi)^3}\Delta_{k'}^{ln}\frac{N_{k'}}{2\kappa_{k'}}+\Delta^l_{\left(k\right)n}k_k\int\frac{d^3k'}{(2\pi)^3}\Delta_{k'}^{jn}k'_lk'_m\frac{N_{k'}}{2\kappa_{k'}}\right]+\left(k\leftrightarrow m\right)
\tea
We filter out a longitudinal term and use again the spherical symmetry to get

\bea
&&\frac{\delta^2\Gamma_Q}{\delta \bar A_j\left(x,t\right)\delta \bar P^{km}\left(y,t'\right)}=-i\frac{\tau}2\theta\left(t-t'\right)\int\frac{d^3k}{(2\pi)^3}e^{i\bm{k(x-y)}}e^{-\kappa_{k}\left(t-t'\right)}(\Delta^j_kk_m+\Delta^j_mk_k)\nn
&&\left[\frac23k^2\int^kdk'\;E[k']+\frac1{15}\int^kdk'\;k'^2E[k']\right]
\tea
$E[k']$ is given in eq. (\ref{KS}).  Since the $k'$ integrals are cut off at $k'\le k$, the first integral is the dominant one. We observe that the integral diverges as $k_c\to 0$, keeping only the most divergent term we get

\be 
\int^kdk'\;E[k']\approx\frac{C_K\epsilon^{2/3}}{k_c^{2/3}}
\te
We finally approximate

\be
\theta\left(t-t'\right)e^{-\kappa_{k}\left(t-t'\right)}\approx \frac{\delta\left(t-t'\right)}{\kappa_{k}}
\te
whereby we find eqs. (\ref{uot}) and (\ref{Lambda1}).

\section{Computation of the Non-Newtonian spectrum}\label{spectcomp}

In this Appendix we give details of the derivation of eq. (\ref{spectrum}). We begin by writing $P[\omega]$ from eq. (\ref{Pomega}) as

\be
P[\omega]=\tau(\omega+i\kappa_k\nu_+)(\omega+i\kappa_k\nu_-)
\label{factor}
\te 
Both $\nu_{\pm}$ are real and positive; when $\tau\to 0$, $\nu_+\to 1$ and $\nu_-\to\infty$; this shows we are dealing with a singular limit. 

The integral over $\omega$ in (\ref{spectrum}) is computed by contour methods and yields

\be 
\int \frac{d\omega}{(2\pi)}\left|G\left[k,\omega\right]\right|^2=\frac{[1+\kappa_k^2\tau^2\nu_+\nu_-]}{2\kappa_k^3\tau^2\nu_+\nu_-(\nu_++\nu_-)}
\te 
Comparing eqs. (\ref{Pomega}) and (\ref{factor}) we find

\bea 
\nu_++\nu_-&=&\frac1{\tau\kappa_k}[1+\tau\zeta(\epsilon k^2)^{1/3}]\nn
\nu_+\nu_-&=&\frac1{\tau\kappa_k}[1-\Lambda\tau\nu k^2]
\tea
To obtain eq. (\ref{spectrum}) we substitute these expressions and develop the result to first order in $\Lambda$.
\section{From MSR to the Fokker-Planck equation for the velocity field}\label{MSR4LET}

While in this paper we have emphasized an approach to turbulence based on finding evolution equations for the velocity correlations, another important approach is based on the evolution of the probability density function (pdf) for the velocity field at a given time \cite{LET}. In this appendix we shall derive this evolution equation from the MSR path integral. Concretely, we show that the Fokker-Planck equation for the velocity field pdf is the Feynman-Kac formula approppriate to the MSR path integral \cite{Schulman}.

Our starting point are the randomly driven NSE

\be
v^i_{,t}+V^i=f^i
\label{LIBNSE}
\te 
where 

\be
V^i=v^j\partial_jv^i+\partial_ip-\nu{\Delta}v^i
\te
$\nu$ is the molecular viscosity, and $p$ is the pressure, which we regard as just a functional of the velocity field at a given time, according to the Coulomb-like formula

\be
p\left({x},t\right)=\int d^3x'\;\frac{\partial_iv^j\left({x}',t\right)\partial_jv^i\left({x}',t\right)}{4\pi\left|{x}-{x'}\right|}
\te
The $f^i$, which also obey $\partial_if^i=0$, are the random forces, which we assume to be Gaussian with zero mean and covariance

\be
\left\langle f^i\left({x},t\right)f^j\left({x'},t'\right)\right\rangle=\delta\left(t-t'\right)N^{ij}\left({x-x'}\right)
\te 
where

\be
N^{ij}\left({x}\right)=\int\frac{d^3k}{\left(2\pi\right)^3}\;e^{i{kx}}\Delta_{{k}}^{ij}N\left(k\right)
\te
where $k=\left|{k}\right|$ and 

\be
\Delta_{{k}}^{ij}=\delta^{ij}-\frac{k^ik^j}{k^2}
\te 
We shall assume the NSE (\ref{LIBNSE}) are the continuum limit of a discrete time evolution

\be
v_{k+1}^i\left(x^j\right)=v_{k}^i\left(x^j\right)+dt\left[-V_{k}^i\left(x^j\right)+f_{k}^i\left(x^j\right)\right]
\label{DNSE}
\te
where $v_k^i\left(x^j\right)=v^i\left(x^j,t_k\right)$, $t_k=kdt$. Under this assumption the NS evolution preserves volume in the space of velocity fields.

Our fundamental hypothesis is that the NSE admit only one solution for given initial conditions and realization of the driving forces, which moreover are regarded as independent of each other. Under these hypothesis the probability for the velocity field $v^i\left(t\right)$ to follow a particular evolution between times $0$ and $t_f$ is

\be
\mathcal{P}\left[v^i\left(t\right)\right]=\int Df F\left[f\right]P\left[v^i\left(0\right),0\right] \delta\left(v^i\left(t\right)-v^i\left[t;v^i\left(0\right),f^i\right]\right)
\te 
Here, $F$ is the noise pdf

\be
F\left[f\right]= e^{-\frac12\int d^3xd^3x'dt\;f^i\left(x,t\right)N^{-1}_{ij}\left(x-x'\right)f^j\left(x',t\right)}
\te
We assume that normalization constants are already included in the integration measure. $P\left[v^i\left(0\right),0\right]$ is the pdf for the velocity field at the initial time $t=0$, and $v^i\left[t;v^i\left(0\right),f^i\right]$ is the unique solution to the NSE for the given initial conditions and random driving. We may use the identity

\be
\delta\left(v^i\left(t\right)-v^i\left[t;v^i\left(0\right),f^i\right]\right)=\left({\rm{Det}}\;\frac{\delta\left[\dot v^i+V^i\right]}{\delta v^j}\right)\delta\left(\dot v^i+V^i-f^i\right)
\label{identity}
\te
Under the discretization rules above the determinant has value $1$. We exponentiate the delta function adding an auxiliary field $A_i$ and integrate over the driving forces to get

\be
\mathcal{P}\left[v^i\left(t\right)\right]=P\left[v^i\left(0\right),0\right]\int DA e^{-\frac12\int d^3xd^3x'dt\;A_i\left(x,t\right)N^{ij}\left(x-x'\right)A_j\left(x',t\right)+i\int d^3xdt\;A_i\left(x,t\right)\left[\dot v^i+V^i\right]}
\te 
We may use this pdf for the space-time velocity field to define a generating function, as it is usually done in the MSR approach, leading eventually to an effective action for the mean velocity field. However, it may be used as well to find the pdf for the instantaneous velocity field, by summing over all histories with a common final configuration

\be
P\left[v_f,t_f\right]=\int_{v^i\left(t_f\right)=v^i_f}Dv\;\mathcal{P}\left[v^i\left(t\right)\right]
\te
To find the evolution equation for this pdf we consider the path integral expression for $P\left[v_f,t_f+dt\right]$. We split the path integral at time $t_f$, where the velocity field takes the value $v^i_m$. Then we obtain

\bea
&&P\left[v_f,t_f+dt\right]=\int Dv_m\int_{v^i\left(t_f+dt\right)=v^i_f;v^i\left(t_f\right)=v^i_m }Dv\;P\left[v_m^i,t_f\right]\nn
&&\int DA e^{-\frac12\int_{t_f}^{t_f+dt} d^3xd^3x'dt\;A_i\left(x,t\right)N^{ij}\left(x-x'\right)A_j\left(x',t\right)+i\int_{t_f}^{t_f+dt} d^3xdt\;A_i\left(x,t\right)\left[\dot v^i+V^i\right]}
\tea 
over the short lapse $dt$, the integral is dominated by the linear interpolation

\be
v^i\left(x,t\right)=v_f^i\left(x\right)-\left(v_f^i-v_m^i\right)\left(1-\frac1{dt}\left(t-t_f\right)\right)
\te
We integrate over the auxiliary field and expand to order $dt$, assuming that $v_f^i-v_m^i$ is effectively restricted to a range $O\left(\sqrt{dt}\right)$. Also, we observe that over the short lapse $dt$, we must fall back to the discrete time dynamics eq. (\ref{DNSE}). Under the discretization rules discussed above, this means the $V^i$ factors are evaluated at $t_f$, that is, they are evaluated at the $v_m^i$ velocity field. If the dynamics is not discretized in this way, we cannot assume that the functional determinant in eq. (\ref{identity}) is a constant, and we must compute it explicitly, see ref. (\cite{Zin93}).
\bea
&&P\left[v_f,t_f+dt\right]=\int Dv_m\;e^{-\frac1{2dt}\int d^3xd^3x'\left(v_f^i-v_m^i\right)\left(x\right)N^{-1}_{ij}\left(x-x'\right)\left(v_f^j-v_m^j\right)\left(x'\right)}\nn
&&\left\{1-\int d^3xd^3x'\left(v_f^i-v_m^i\right)\left(x\right)N^{-1}_{ij}\left(x-x'\right)V^j\left[v_f\right]\left(x'\right)\right.\nn
&+&dt\int d^3xd^3x'd^3y\left(v_f^i-v_m^i\right)\left(x\right)N^{-1}_{ij}\left(x-x'\right)\frac{\delta V^j\left(x'\right)}{\delta v^k\left(y\right)}\left(v_f^k-v_m^k\right)\left(y\right)\nn
&-&\frac12dt\int d^3xd^3x'V^i\left[v_f\right]\left(x\right)N^{-1}_{ij}\left(x-x'\right)V^j\left[v_f\right]\left(x'\right)\nn
&+&\left.\frac12\left[\int d^3xd^3x'\left(v_f^i-v_m^i\right)\left(x\right)N^{-1}_{ij}\left(x-x'\right)V^j\left[v_f\right]\left(x'\right)\right]^2\right\}\nn
&&\left\{P\left[v_f,t_f\right]-\int d^3x\frac{\delta P\left[v_f,t_f\right] }{\delta v_f^i\left(x\right)}\left(v_f^i-v_m^i\right)\left(x\right)\right.\nn
&+&\left.\frac12\int d^3xd^3x'\frac{\delta^2 P\left[v_f,t_f\right] }{\delta v_f^i\left(x\right)\delta v_f^j\left(x'\right)}\left(v_f^i-v_m^i\right)\left(x\right)\left(v_f^j-v_m^j\right)\left(x'\right)\right\}
\tea
We finally compute the Gaussian averages from the covariance

\be
\left\langle \left(v_f^i-v_m^i\right)\left(x\right)\left(v_f^j-v_m^j\right)\left(x'\right)\right\rangle=dt N^{ij}\left(x-x'\right)
\te
and take the $dt\to 0$ limit, whereby

\bea
\frac{\partial P\left[v_f,t_f\right] }{\partial t_f}&=&\int d^3x\left\{\frac{\delta V^i\left(x\right)}{\delta v^i\left(x\right)}P\left[v_f,t_f\right]+V^i\left(x\right)\frac{\delta P\left[v_f,t_f\right]}{\delta v^i\left(x\right)}\right\}\nn
&+&\frac12\int d^3xd^3x'\;N^{ij}\left(x-x'\right)\frac{\delta^2 P\left[v_f,t_f\right] }{\delta v_f^i\left(x\right)\delta v_f^j\left(x'\right)}
\tea
Which is the required equation.

\end{document}